\documentclass[usenatbib]{mn2e}
\usepackage{graphicx,fixltx2e}
\usepackage{wasysym}

\def\rpd{\hbox{rad\,d$^{-1}$}}

\def\chisq{\hbox{$\chi^2$}}
\def\chisqr{\hbox{$\chi^2_{\rm r}$}}
\def\mSun{\hbox{${\rm M}_{\odot}$}}

\def\rSun{\hbox{${\rm R}_{\odot}$}}

\def\sn{\hbox{S/N}}
\def\vrad{\hbox{$v_{\rm rad}$}}
\def\ms{\hbox{m\,s$^{-1}$}}

\def\kms{\hbox{km\,s$^{-1}$}}
\def\vsini{\hbox{$v \sin i$}}

\def\ptt{\hbox{$10^{-4} I_{\rm c}$}}

\def\degr{\hbox{$^\circ$}}

\def\omeq{\hbox{$\Omega_{\rm eq}$}}
\def\dom{\hbox{$d\Omega$}}

\newcommand{\caii}{\hbox{Ca$\;${\sc ii}}}

\begin{document}

\title[]
{Searching for Star-Planet interactions within the magnetosphere of HD 189733
 }

\makeatletter

\def\newauthor{%
  \end{author@tabular}\par
  \begin{author@tabular}[t]{@{}l@{}}}
\makeatother
\author[R.~Fares et al.]{\vspace{1.7mm} 
R.~Fares$^{1,2}$\thanks{E-mail:
rim.fares@ast.obs-mip.fr (RF);
donati@ast.obs-mip.fr (J-FD);
claire.moutou@oamp.fr (CM); 
mmj@st-andrews.ac.uk (MMJ);
griessmeier@astron.nl (J-MG);
philippe.zarka@obspm.fr (PZ);
shkolnik@dtm.ciw.edu (ES);
david.bohlender@nrc-cnrc.gc.ca (DB); 
claude.catala@obspm.fr (CC); 
acc4@st-andrews.ac.uk (ACC)},
J.-F.~Donati$^1$, C.~Moutou$^2$, M.M.~Jardine$^3$, J.-M. Grie{\ss}meier$^4$, P. Zarka$^5$\\ 
\vspace{1.7mm}
{\hspace{-1.5mm}\LARGE\rm
E.~L.~Shkolnik$^6$, D.~Bohlender$^7$, C.~Catala$^5$ , A.C.~Cameron$^3$}\\
$^1$ LATT--UMR 5572, CNRS \& Univ.\ P.~Sabatier, 14 Av.\ E.~Belin, F--31400 Toulouse, France \\
$^2$ LAM--UMR 6110, CNRS \& Univ.\ de Provence, 38 rue Fr\'ederic Juliot-Curie, F--13013 Marseille, France \\
$^3$ School of Physics and Astronomy, Univ.\ of St~Andrews, St~Andrews, Scotland KY16 9SS, UK \\
$^4$ Netherlands Institute for Radio Astronomy, Postbus 2, 7990 AA, Dwingeloo, The Netherlands\\
$^5$ LESIA--UMR 8109, CNRS \& Univ.\ Paris VII, 5 Place Janssen, F--92195 Meudon Cedex, France \\
$^6$ Department of Terrestrial Magnetism, Carnegie Institution of Washington,  5241 Broad Branch Road, NW, Washington, DC 20015-130 , USA \\ 
$^7$ HIA/NRC, 5071 West Saanich Road, Victoria, BC V9E 2E7, Canada \\ 
}
\date{accepted, 11 March 2010}
\maketitle

\begin{abstract}

HD~189733 is a K2 dwarf, orbited by a giant planet at 8.8 stellar radii. In order to study magnetospheric interactions between the star and the planet, we explore the large-scale magnetic field and activity of the host star. 

We collected spectra using the ESPaDOnS and the NARVAL spectropolarimeters, installed at the 3.6-m Canada-France-Hawaii telescope and the 2-m Telescope Bernard Lyot at Pic du Midi, during two monitoring campaigns (June~2007 and July~2008).

HD~189733 has a mainly toroidal surface magnetic field, having a strength that reaches up to 40 G. The star is differentially rotating, with latitudinal angular velocity shear of $\dom = 0.146 \pm 0.049$~\rpd, corresponding to equatorial and polar periods of $11.94 \pm 0.16 $~d and $16.53 \pm 2.43$~d respectively. The study of the stellar activity shows that it is modulated mainly by the stellar rotation (rather than by the orbital period or the beat period between the stellar rotation and the orbital periods). We report no clear evidence of magnetospheric interactions between the star and the planet.

We also extrapolated the field in the stellar corona and calculated the planetary radio emission expected for HD~189733b given the reconstructed field topology. The radio flux we predict in the framework of this model is time variable and potentially detectable with LOFAR.

\end{abstract}


\section{Introduction}

\label{sec:intro}

Magnetic fields are present at different scales in the universe, from planets, to stars, galaxies and galaxy clusters. Thanks to new high-resolution spectropolarimeters, we are able to study the large scale magnetic field of stars, trying to understand its origin, but also its implication in different stellar phenomena (stellar wind, magnetic braking, stellar cycles, ...). In Hot Jupiter (HJ) systems (giant planets orbiting close to their parent stars, i.e. with semi-major axis of the planet lower than 0.1 AU), the study of the stellar magnetic field is important to understand Star-Planet interactions (SPI), and their effects on the evolution and properties of the system. 

SPI can be of two types: magnetospheric (e.g. caused by reconnections between the stellar and the planetary magnetic fields), or tidal (resulting from the proximity and masses of the two bodies). \cite{cuntz00} suggested that such interactions may enhance the stellar activity. Studying the activity of HJ hosting stars, \cite{shk03,shk05,shk08} concluded that not all the observed systems show hints of interactions and that for a single system with interactions, those interactions may be not observable during some observing epochs, yet observable at other times.

Different theoretical scenarios of magnetospheric interactions were proposed. \cite{preusse06} described SPI by adopting the Alfv\'en wind model, \cite{lanza08} considered a non-potential magnetic field configuration for the closed corona of the star. To explain the 'on-off' nature of SPI, \cite{cranmer07} studied their signatures over many orbital cycles considering a cyclic stellar magnetic field. These signatures do not repeat exactly from orbit to orbit, nor from epoch to epoch. SPI is an intermittent phenomenon depending strongly on the configuration of the stellar field. 

The study of the stellar magnetic field is therefore important to understand SPI. We have started an observing program aimed at detecting and modelling the magnetic field of HJ hosting stars. Eleven systems are observed, having different stellar and planetary parameters. In this paper, we present the results for HD~189733. This system is interesting to study SPI, with a short orbital period different from the stellar rotation period. 

The properties of HD~189733 are listed in section \ref{sec:star}. In sections \ref{sec:obs} and \ref{sec:mod}, we will present our data, data analysis and magnetic modelling of the star. Results of its magnetic topology and differential rotation will also be presented. Stellar activity will be studied in section \ref{sec:Activity}. From magnetic maps of the stars obtained in section \ref{sec:mod}, we will extrapolate the magnetic field in the stellar corona (section \ref{sec:extrapolation}), deduce the expected planetary radio emission (section \ref{sec:radio}), and end with our conclusions (section \ref{sec:conclusions}).


\section{Properties of HD~189733}
\label{sec:star}
%

HD~189733 is a well-known planet hosting star, discovered in 2005 by \cite{bouchy05}. The K2V star is bright, nearby and active (V=7.7, $T_{\rm eff}= 5050\pm50~\rm{K}$, $\rm [Fe/H]=-0.03\pm0.04$~\citealt{bouchy05}, d=19.3~pc). It has a mass of $0.82\pm0.03~\mSun$~\citep{bouchy05} and a radius $\rm R_{\star}=0.76\pm0.01~\rSun$~\citep{winn07}. \cite{winn06} measured \vsini~($2.97\pm0.22~\kms$). The rotation period of the star was measured by different teams using photometry. \cite{hebrard06} found it to be about 11.8~d, while \cite{winn07} found quasi-periodic flux variations of 13.4~d period, which they attributed to the stellar rotation. These different values may be due to temporal variations in spots coupled to differential rotation over stellar latitudes. 

As the planet transits HD~189733, all the planetary parameters are well constrained. A recent review of the planetary orbit was done by \cite{boisse09}. The inclination angle of the orbit is of $85.76\degr\pm0.29\degr$, the planet has a mass of $1.13\pm0.03~\rm{M_{\jupiter}}$~and a radius of $1.154\pm0.032~\rm{R_{\jupiter}}$~. It orbits the star every $2.2185733\pm0.0000019$~d on a circular orbit. The semi-major axis of the orbit is of $0.031\pm0.001~\rm{AU}$~(\cite{boisse09} and references therein). \cite{triaud09} measured the projected spin-orbit misalignment angle and found it to be $0.85\degr^{+0.32}_{-0.28}$, i.e., both orbital and stellar rotation axes are almost aligned.


\section{Observations}
\label{sec:obs}
%
 
We observed HD~189733 in June 2007 and July 2008 using ESPaDOnS and NARVAL spectropolarimeters, the former installed at the 3.6-m Canada-France-Hawaii Telescope (CFHT) and the latter at the 2-m Telescope Bernard Lyot (TBL) at Pic du Midi. These spectropolarimeters provide spectra with a resolution of 65000 which span the whole optical domain (370 to 1000~nm). Each polarization spectrum is extracted from a sequence of four subexposures, taken in different configurations of the polarimeter waveplates, in order to perform a full circular polarization analysis.

The data were reduced using a fully automatic reduction tool, called Libre-Esprit, installed at CFHT and TBL for the use of observers \citep{donati97}. The spectra have error bars at each wavelength pixel and are normalized to a unit continuum. Their wavelength scale refers to the heliocentric rest frame. They are automatically corrected from spectral shifts resulting from instrumental effects (e.g. mechanical flexures, temperature or pressure variations) using telluric lines as a reference. Though not perfect, this procedure allows spectra to be secured with a radial velocity (RV) precision of better than 30~\ms~\citep{moutou07,morin08}. 

We collected 20 spectra in June/July~2007 using both ESPaDOnS and NARVAL. These data covers about 2 stellar rotations (25 nights); NARVAL data cover roughly the first rotation cycle, and ESPaDOnS data sample well the second rotational cycle. In 2008, we only used NARVAL. Our data are spread over 14 nights, from 10 to 24~July. This second data set covers only slightly more than one rotation cycle. At around 700~nm, the S/N ratio of the spectra varies between 220 and 940 for 2007 spectra, and from 390 to 720 for 2008 spectra (per 2.6~\kms\ velocity bin). We also use the 2006 data published in \cite{moutou07} merged in a single data set (as opposed to \cite{moutou07} who analyzed separately the June~2006 and August~2006 data). The phase coverage is poor for this epoch. The observations, separated by 52~days, overlap in terms of rotation cycle. The complete log of the 2007 and 2008 observations is given in Table \ref{tab:logobs} \citep[refer to][for information about the 2006 data]{moutou07}.

The rotational and orbital phases, denoted $\rm{E_{\rm Rot}}$ and $\rm{E_{\rm Orb}}$, were computed using the two ephemerides:
\begin{eqnarray}
T_0 &=& \mbox{HJD~}2,453,629.389 + 2.218574~\rm{E_{\rm Orb}} \nonumber\\
T_0 &=& \mbox{HJD~}2,453,629.389 + 12~\rm{E_{\rm Rot}}
\label{eq:eph}
\end{eqnarray}
The first ephemeris is that of \cite{moutou07}. For the ephemeris giving the rotation phase, we use a rotation period of 12~d, identified as the equatorial rotation period (see section \ref{sec:mod}).  

\begin{table*}
\caption[]{Journal of June~2007 and July~2008 observations. Columns 1--11 sequentially list the UT date, the instrument used, the heliocentric Julian date and UT time (both at mid-exposure), the complete exposure time, the peak signal to noise ratio (per 2.6~\kms\ velocity bin) of each observation (around 700~nm), the rotational and orbital cycles (using the ephemeris given by Eq.~\ref{eq:eph}), the radial velocity (RV) associated with each exposure, the rms noise level (relative to the unpolarized continuum level
$I_{\rm c}$ and per 1.8~\kms\ velocity bin) in the circular polarization profile produced by Least-Squares Deconvolution (LSD) and the longitudinal magnetic field.  }
\begin{tabular}{ccccccccccc}
\hline
\hline
Date (UT) & Instrument & HJD          & UT      & $t_{\rm exp}$  & \sn & Rot. Cycle & Orb. Cycle & \vrad & $\sigma_{\rm LSD}$&$B_{l}$\\
&    & (2,454,000+) & (h:m:s) & (s) & & (52+) & (284+)&(\kms)&(\ptt)&(G)  \\
\hline
09 June 07  & NARVAL & 260.532660 & 0:42:06 & 4$\times$900& 210 & 0.5953 & 0.4817 &-2.213 & 1.59 & $-3.5\pm3.2$\\
13 June 07  & NARVAL & 264.561950 & 1:23:52 & 4$\times$900& 390 & 0.9311 & 2.2978 &-2.382 & 0.84 & $-0.2\pm1.7$\\
18 June 07  & NARVAL & 269.559040 & 1:19:11 & 4$\times$900& 330 & 1.3475 & 4.5502 &-2.152 & 1.01 & $-1.0\pm2.0$\\
19 June 07  & NARVAL & 271.479030 & 23:23:48 & 4$\times$900& 360 & 1.5075 & 5.4156 &-2.299 & 0.97 & $-1.3\pm1.9$\\
23 June 07  & ESPaDOnS& 275.004850 & 12:00:39 & 4$\times$900& 920 & 1.8013 & 7.0049 &-2.294 & 0.34 & $-2.1\pm0.7$\\
23 June 07  & ESPaDOnS& 275.121020 & 14:47:56 & 4$\times$800 & 870 & 1.8110 & 7.0572 &-2.340 & 0.36 & $-0.6\pm0.7$\\
26 June 07  & ESPaDOnS& 277.847340 & 8:13:36 & 4$\times$900 & 740 & 2.0382 & 8.2861 &-2.416 & 0.45 & $-3.2\pm0.9$\\
26 June 07  & ESPaDOnS& 278.117560 & 14:42:43 & 4$\times$900 & 870 & 2.0607 & 8.4079 &-2.352 & 0.36 & $-3.3\pm0.7$\\
27 June 07  & ESPaDOnS& 278.841160 & 8:04:38 & 4$\times$800& 910 & 2.1210 & 8.7341 &-1.997 & 0.36 & $-6.2\pm0.7$\\
28 June 07  & ESPaDOnS& 279.840490 & 8:03:35 & 4$\times$900 & 810 & 2.2043 & 9.1845 &-2.390 & 0.40 & $-2.9\pm0.8$\\
30 June 07  & ESPaDOnS& 281.863400 & 8:36:26 & 600+140 & 270 & 2.3729 & 10.0963 &-2.325 & 1.19 & $-2.0\pm2.3$\\
01 July 07  & ESPaDOnS& 282.834040 & 7:54:05 & 4$\times$700& 770 & 2.4538 & 10.5338 &-2.164 & 0.44 & $-2.2\pm0.9$\\
01 July 07  & ESPaDOnS& 283.120060 & 14:45:56 & 4$\times$800 & 690 & 2.4776 & 10.6627 &-2.039 & 0.49 & $-3.7\pm1.0$\\
02 July 07  & ESPaDOnS& 283.949800 & 10:40:42 & 4$\times$600& 590 & 2.5467 & 11.0367 &-2.286 & 0.57 & $-5.1\pm1.1$\\
02 July 07  & ESPaDOnS& 284.131990 & 15:03:02 & 4$\times$600& 640 & 2.5619 & 11.1188 &-2.331 & 0.52 & $-3.9\pm1.0$\\
03 July 07  & ESPaDOnS& 284.975470 & 11:17:36 & 4$\times$600 & 760 & 2.6322 & 11.4990 &-2.262 & 0.42 & $-3.5\pm0.8$\\
03 July 07  & ESPaDOnS& 285.112190 & 14:34:28 & 4$\times$600 & 660 & 2.6436 & 11.5607 &-2.149 & 0.50 & $-2.9\pm1.0$\\
04 July 07  & ESPaDOnS& 285.842200 & 8:05:38 & 4$\times$600 & 510 & 2.7044 & 11.8897 &-2.089 & 0.65 & $-5.0\pm1.3$\\
04 July 07  & ESPaDOnS& 286.014210 & 12:13:19 & 4$\times$600 & 490 & 2.7188 & 11.9672 &-2.167 & 0.68 & $-3.6\pm1.3$\\
04 July 07  & ESPaDOnS& 286.087050 & 13:58:13 & 4$\times$900 & 450 & 2.7248 & 12.0001 &-2.184 & 0.74 & $-3.9\pm1.5$\\
\hline
10 July 08  & NARVAL & 658.455050 & 22:47:44 & 4$\times$900& 570 & 33.7555 & 179.8412 &-1.977 & 0.55 & $-3.7\pm1.1$\\
11 July 08  & NARVAL & 658.603660 & 2:21:43 & 4$\times$900& 590 & 33.7679 & 179.9082 &-2.014 & 0.56 & $-2.7\pm1.1$\\
11 July 08  & NARVAL & 658.636700 & 3:0:9:18 & 2$\times$900& 360 & 33.7706 & 179.9231 &-2.025 & 0.90 & $-4.4\pm1.8$\\
14 July 08  & NARVAL & 662.448330 & 22:37:52 & 4$\times$900& 440 & 34.0883 & 181.6411 &-1.970 & 0.74 & $-3.8\pm1.5$\\
15 July 08  & NARVAL & 662.589210 & 2:00:44 & 4$\times$900& 480 & 34.1000 & 181.7046 &-1.942 & 0.69 & $-1.4\pm1.4$\\
15 July 08  & NARVAL & 662.633270 & 3:04:11 & 4$\times$900& 510 & 34.1037 & 181.7245 &-1.939 & 0.65 & $-1.8\pm1.3$\\
15 July 08  & NARVAL & 663.446250 & 22:34:51 & 4$\times$900& 360 & 34.1714 & 182.0909 &-2.230 & 0.95 & $1.8\pm1.9$\\
16 July 08  & NARVAL & 663.587830 & 1:58:42 & 4$\times$900& 440 & 34.1832 & 182.1548 &-2.284 & 0.78 & $3.3\pm1.5$\\
16 July 08  & NARVAL & 663.631880 & 3:02:09 & 4$\times$900& 390 & 34.1869 & 182.1746 &-2.298 & 0.90 & $4.2\pm1.8$\\
17 July 08  & NARVAL & 665.486420 & 23:32:37 & 4$\times$900& 730 & 34.3415 & 183.0105 &-2.129 & 0.42 & $-0.3\pm0.8$\\
18 July 08  & NARVAL & 665.587170 & 1:57:42 & 4$\times$900& 720 & 34.3498 & 183.0559 &-2.168 & 0.43 & $-1.1\pm0.9$\\
18 July 08  & NARVAL & 666.466700 & 23:04:11 & 4$\times$900& 660 & 34.4231 & 183.4524 &-2.193 & 0.48 & $0.0\pm0.9$\\
19 July 08  & NARVAL & 666.569380 & 1:32:03 & 4$\times$900& 630 & 34.4317 & 183.4987 &-2.122 & 0.49 & $-0.3\pm1.0$\\
19 July 08  & NARVAL & 666.657390 & 3:38:47 & 2$\times$900& 420 & 34.4390 & 183.5383 &-2.063 & 0.70 & $1.2\pm1.4$\\
20 July 08  & NARVAL & 668.491560 & 23:39:55 & 4$\times$900& 690 & 34.5919 & 184.3651 &-2.263 & 0.47 & $-8.8\pm0.9$\\
21 July 08  & NARVAL & 668.590150 & 2:01:53 & 4$\times$900& 680 & 34.6001 & 184.4095 &-2.224 & 0.46 & $-7.3\pm0.9$\\
21 July 08  & NARVAL & 669.465410 & 23:02:15 & 4$\times$900& 430 & 34.6730 & 184.8040 &-1.939 & 0.78 & $-5.0\pm1.6$\\
22 July 08  & NARVAL & 669.565650 & 1:26:35 & 4$\times$900& 510 & 34.6814 & 184.8492 &-1.952 & 0.63 & $-1.5\pm1.3$\\
22 July 08  & NARVAL & 669.659660 & 3:41:57 & 2$\times$900& 420 & 34.6892 & 184.8916 &-1.968 & 0.72 & $-2.4\pm1.4$\\
22 July 08  & NARVAL & 670.469810 & 23:08:33 & 4$\times$900& 580 & 34.7567 & 185.2567 &-2.337 & 0.55 & $-2.2\pm1.1$\\
23 July 08  & NARVAL & 670.570050 & 1:32:54 & 4$\times$900& 670 & 34.7651 & 185.3019 &-2.323 & 0.47 & $0.9\pm0.9$\\
23 July 08  & NARVAL & 670.651980 & 3:30:52 & 2$\times$800& 430 & 34.7719 & 185.3389 &-2.305 & 0.69 & $1.1\pm1.4$\\
24 July 08  & NARVAL & 672.478300 & 23:20:44 & 4$\times$900& 680 & 34.9241 & 186.1621 &-2.266 & 0.47 & $-4.2\pm0.9$\\
25 July 08  & NARVAL & 672.577800 & 1:44:01 & 4$\times$900& 720 & 34.9324 & 186.2069 &-2.277 & 0.44 & $-4.9\pm0.9$\\
\hline
\end{tabular}
\label{tab:logobs}
\end{table*}

To improve the S/N ratio of our data and extract the polarization from many lines simultaneously, we used Least-Squares Deconvolution (LSD) which assumes that all lines more or less repeat the same polarization information. In practice, it consists of deconvolving the observed spectra by a line mask, computed using Kurucz's lists of atomic line parameters (Kurucz CD-Rom 18) and a Kurucz model atmosphere with solar abundances, temperature and logarithmic gravity (in \hbox{cm\,s$^{-2}$}) set to $5000$~K and 4.0 respectively. The line mask includes the moderate to strong lines present in the optical domain (those featuring central depths larger than 40\% of the local continuum, before any macroturbulent or rotational broadening, about 4,000 lines throughout the whole spectral range) but excludes the strongest, broadest features, such as Balmer lines, whose Zeeman signature is strongly smeared out compared to those of narrow lines. In addition to the intensity and polarization profiles, LSD produces a null profile (labelled N) that should contain no polarization; this helps to confirm that the detected polarization is real and not due to spurious instrumental or reduction effects \citep{donati97}. The multiplex gain provided by LSD in V and N spectra is of the order of 30 with regard to a single line with average magnetic sensitivity, implying noise levels as low as 35 parts per million (ppm). A typical set of LSD profiles is shown in Fig.~\ref{fig:profilLSD} for 20~July~2008.
\begin{figure}
\includegraphics[height=.25\textheight]{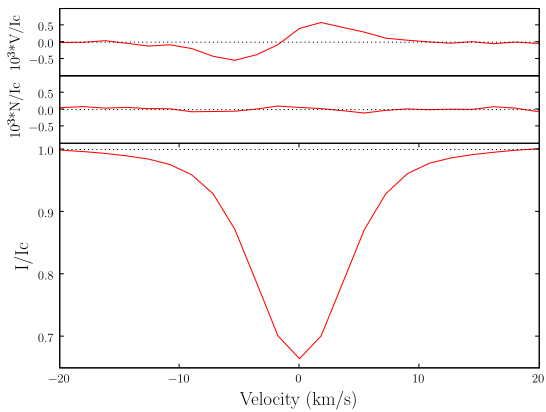}
\caption[]{Typical LSD profiles shown for 20~July~2008.}
\label{fig:profilLSD}
\end{figure}

The radial velocity (RV) of the star is obtained by fitting each Stokes I profile with a Gaussian profile. The RV values we obtain are listed in Table \ref{tab:logobs} and are in good agreement with the expectations, when using the orbital solution of \citet{boisse09} once the data are globally shifted by $-0.06$~\kms\ and $-0.15$~\kms\ for June 2007 and July 2008 respectively (see Fig.~\ref{fig:vrad}). A similar offset was reported by \cite{moutou07}. These variable RV shifts may be due to stellar variability, to the presence of a stellar companion \citep{bakos06b}, or to a yet unknown cause. The wavelength calibration being done relative to the telluric lines (giving a precision of about 30~\ms), we believe it is very unlikely that these RV shifts could be due to any instrumental effect.

For the magnetic analysis in the rest of the paper, the spectra used are the LSD ones, corrected for the orbital motion of the planet.
\begin{figure}
\includegraphics[height=.25\textheight]{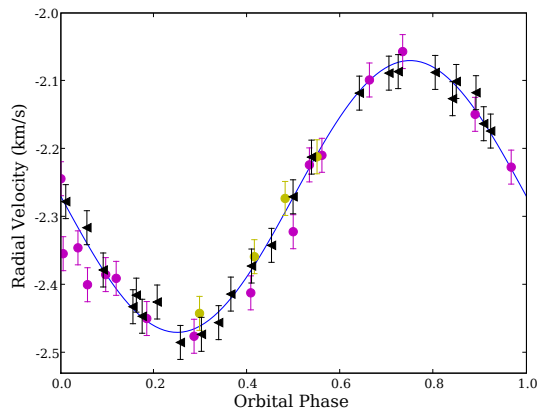}
\caption[]{Radial velocities of HD~189733 derived from June 2007 (NARVAL: yellow dots and ESPaDOnS: magenta dots) and July 2008 (black triangles) spectra as a function of orbital phase, with their error bars ($20 - 30$~\ms) shifted by 0.06~\kms\ and 0.15~\kms\ respectively (see text for more details). The radial velocity model plotted here (blue full line) corresponds to the orbital solution of \cite{boisse09}. }
\label{fig:vrad}
\end{figure}

\section{Magnetic imaging}
\label{sec:mod}
\subsection{Model description and imaging method}

To reconstruct the magnetic maps of HD~189733, we use the Zeeman Doppler Imaging code \citep{donati97} in its latest version. ZDI is a tomographic imaging technique, it inverts series of circular polarization Stokes V profiles into the parent magnetic topology, i.e. the distribution of magnetic fluxes and orientations. Since the problem is ill-posed, ZDI uses the principles of maximum entropy to retrieve the simplest image compatible with the data. The latest version of the code describes the field by its radial poloidal, non-radial poloidal and toroidal components \citep{Chand61}, all  expressed in terms of spherical harmonics expansions. This has the advantage that both simple and complex topologies can be reconstructed reliably \citep{donati01}. Moreover, the energy of the poloidal and toroidal components, or of axisymmetric and non-axisymmetric modes can be estimated directly from the coefficients of the spherical harmonics expansion. Given the small value of \vsini, the resolution at the surface of the star is limited; we therefore truncate the spherical harmonics expansion to the five lowest terms only, i.e., $l < 5$.

Reconstruction proceeds by iteratively comparing the reconstructed profiles to the observed one, until they match within the error bars (i.e. reduced chi-square $\chisqr  \sim 1$). Practically, the star is divided into 9000 grid cells of similar area. The contribution of each grid cell to the reconstructed Stokes profiles is calculated, given the RV of the cell, the field strength and orientation in it, the location and projected area. Summing the contributions of all the grid cells yields the synthetic profile.

For each grid cell, the local unpolarized Stokes I is modelled by a Gaussian with a full width at half-maximum (FWHM) of 7~\kms, central rest wavelength of 500~nm and effective Land\'e factor of $1.25$. The Stokes V profile is calculated assuming the weak-field approximation, i.e. the V profile is proportional to the line-of-sight projected component of the field (called longitudinal field and denoted $B_{l}$), as well as the first derivative of the local I profile. The inclination angle of the rotation axis of the star with respect to the line-of-sight is $\sim~85\degr$ (see section \ref{sec:star}). 
 
For a differentially rotating star, magnetic regions located at different latitudes have different angular velocities. As in \citet{fares09}, we consider that the rotation at the surface of the star follows $\Omega(\theta) = \omeq - \dom  \sin^2(\theta)$, where $\Omega(\theta)$ and $\omeq$ are respectively the angular velocities at a latitude $\theta$~and at the equator, and $\dom$~is the difference in rotation rate between the pole and the equator. The position of the grid cells are calculated by this law relative to the median observing epoch. Measuring the recurrence rate of the signatures from magnetic regions located at various latitudes gives access to the amount of surface shear. In practice, for each pair of (\omeq, \dom) in acceptable range of values, we reconstruct a magnetic image at a given information content (constant magnetic energy) from the observed profiles and get the \chisqr~of the reconstruction procedure. Fitting a paraboloid to the \chisqr~values we obtain by this process gives the optimum differential rotation (DR) parameters of the star.

\subsection{Results}
\subsubsection{Differential rotation}

\begin{figure}
\includegraphics[scale=0.35, angle=-90]{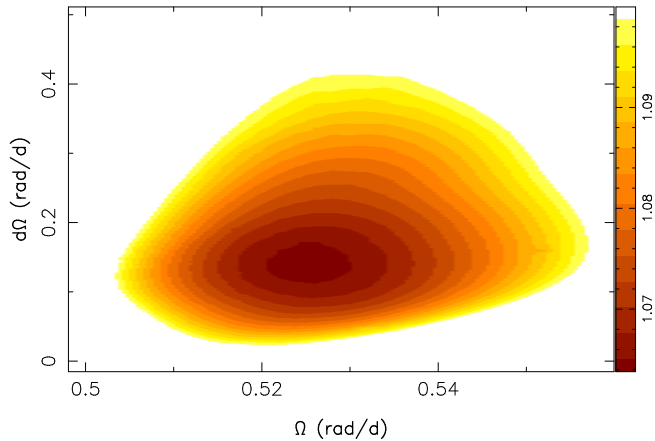}
\caption[]{Variations of \chisqr~as a function of \omeq~and \dom~, derived from the modelling of the Stokes V data set for June 2007. The outer colour contour corresponds to a 3.4\%~increase in the \chisqr~, and traces a 3 $\sigma$ interval for both parameters taken as a pair.}
\label{fig:DR}
\end{figure}

Including DR for the reconstruction of June~2007 map is necessary to fit the observed Stokes profiles almost down to the noise level. We applied our method to measure the DR for this data set. The \chisqr~values as a function of \omeq~and \dom~form a well defined paraboloid shown in Fig. \ref{fig:DR}.  The optimal DR parameters, for which \chisqr~is minimal, are respectively $\omeq = 0.526 \pm 0.007 $~\rpd~and $\dom = 0.146 \pm 0.049$~\rpd~. The equator of HD~189733 rotates in $11.94 \pm 0.16 $~d, while its pole has a slower rotation of $16.53 \pm 2.43$~d. This corresponds to a time for the equator to lap the pole by one complete rotation cycle of $43.06\pm14.44$~d (lap time).

For July~2008, we are not able to measure DR, the \chisqr~map featuring no well defined minimum (over the range of parameters we explored). Our data have a lower S/N ratio than those of June 2007, they cover slightly more than an equatorial rotation period; they do not contain all the information needed for measuring the recurrence rate of the signatures of high latitude magnetic features. For this epoch, the DR parameters we include in our reconstruction procedure are those of June 2007 (for the same \chisqr~value), supposing that the DR did not change between the two epochs.

For the 2006 data, we are not able to measure the DR given the very small phase interval covered both in June and August, the resulting \chisq~map yielding again a very chaotic surface. Nevertheless, adding the differential parameters of June 2007 in our reconstruction gives a better fit to the data (\chisqr~$\sim1.26$ as opposed to 3.9 for solid-body rotation with a rotation period of 12~d). 

\subsubsection{Magnetic maps}

For June~2007, the reconstructed Stokes V profiles including the DR values we obtained are shown in Fig. \ref{fig:profils}, the data being fitted for a \chisqr~value slightly larger than 1. HD~189733 has an average surface magnetic field of $22~\rm{G}$. The field has a predominant toroidal component, contributing 57\% of the total magnetic energy. The poloidal component of the field is mainly non-axisymmetric, 67\%~ of its energy being in modes with $m>l/2$. The quadrupole, octupole and higher orders contribute almost equally to the poloidal energy. Orders with $l>3$ contains 30\%~of the poloidal energy, while 70\%~is in the lower orders.

\begin{figure*}
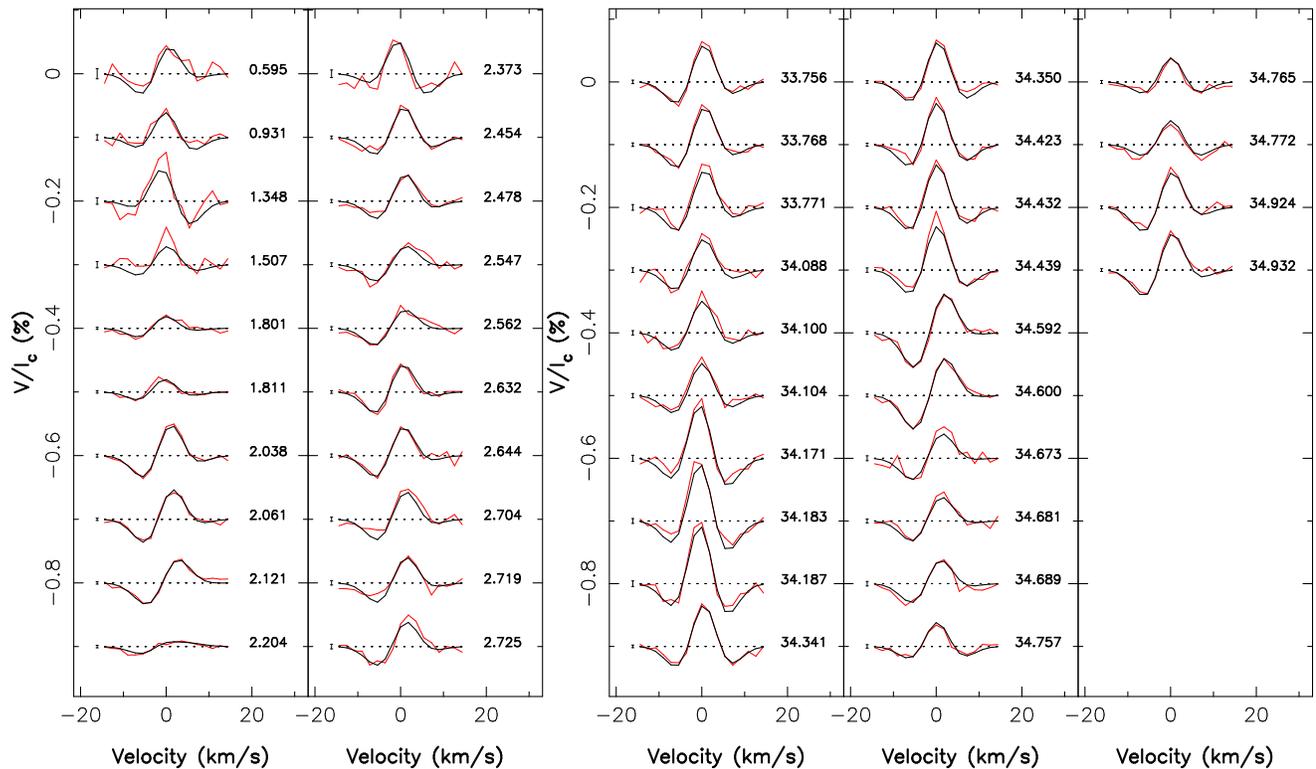

\center{\hbox{\includegraphics[scale=0.55,angle=-90]{ProfilVJun07.ps}\includegraphics[scale=0.55,angle=-90]{ProfilVJul08.ps}}}
\caption[]{Maximum-entropy fits (thin red line) to the observed Stokes $V$~LSD 
profiles (thick black line) of HD~189733 for 2007 June (left) and 2008 July (right).  The rotational cycle of each observation 
(as listed in Table \ref{tab:logobs}) and 1$\sigma$ error bars are also shown next to each profile.  }
\label{fig:profils}
\end{figure*}

\begin{figure*}
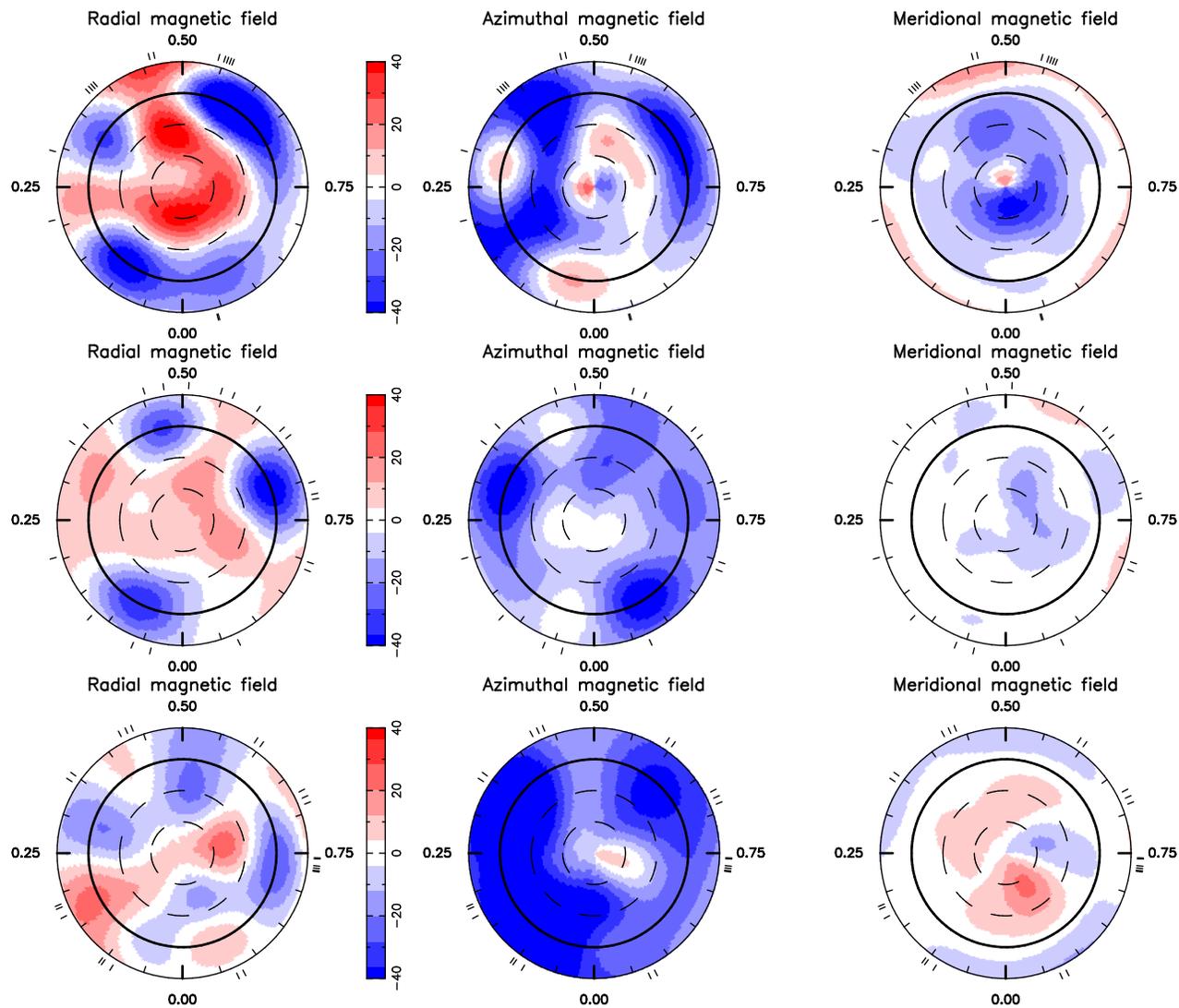

\center{\hbox{\hspace{12mm}\includegraphics[scale=0.7]{MapSum06.ps}\hspace{6mm}}
	\hbox{\hspace{12mm}\includegraphics[scale=0.7]{MapJun07.ps}\hspace{6mm}}
       \hbox{\hspace{12mm}\includegraphics[scale=0.7]{MapJul08.ps}\hspace{6mm}}} 

\caption[]{Maximum-entropy reconstructions of the large-scale magnetic topology of
HD~189733 as derived from our 2006 (top row), 2007 June (middle row) and 2008 July (bottom row) data sets. The radial, azimuthal and meridional components of the field (with magnetic flux values labelled in G).
The star is shown in flattened polar projection down to latitudes of $-30\degr$, with the equator depicted as a bold circle and parallels as dashed circles. Radial ticks around each plot indicate rotational phases of observations. }
\label{fig:maps}
\end{figure*}

In July~2008, the field has a greater average value of $36~\rm{G}$. Its strength reaches up to $40~\rm{G}$ in some magnetic region (see Fig. \ref{fig:maps}). The toroidal component contributes 77\% of the total energy. While the poloidal field is still mainly non-axisymmetric, its dipolar component has a stronger contribution for this epoch. Orders with $l>3$ still have a contribution of about 32\%~to the poloidal energy.

\begin{table}
\caption[]{Average magnetic field on the surface of the star, percentage of the toroidal energy relative
to the total one, percentage of the energy contained in the axisymmetric modes of the poloidal
component for each epoch of observation.}
\begin{tabular}{c|c|c|c|}
\hline
Epoch & B (G)&\%~ of E$_{\rm tor}$&  Axisymmetric modes  \\ 
&&&\%~ of poloidal \\
\hline
2006 & 33&33 & 56 \\ 
\hline
June 2007 & 22&57 & 26  \\  
\hline
July 2008 & 36&77 & 17\\  
\hline
\end{tabular}
\label{tab:prop}
\end{table}

When merging June and August~2006 in one map, the poloidal field contributes 67\%~to the total energy. The average value of the field is of $33~\rm{G}$. The properties of the magnetic field are listed in Table \ref{tab:prop}.

For the three epochs, we do not observe a global change in the magnetic polarity. The radial field shows positive magnetic regions covering the pole and a concentration of the magnetic regions around the equator for all epochs. A drop in the poloidal energy is observed between 2006 and 2007, as well as between 2007 and 2008 (by a factor of 1.6 and 1.9 respectively). The topology of the field also changes over two years, with the toroidal component strengthening significantly at the expense of the poloidal component.

\section{Activity indicators}
\label{sec:Activity}

The \caii~H~\&~K\ and H$\alpha$ lines are tracers of the stellar chromospheric activity. We studied the variability of the residual emission in those activity proxies for our data.
For each tracer, we calculated a mean profile per run (see Fig.~\ref{fig:meanCa} in the particular case of July~2008), then subtract it from each spectrum. The residual emission profiles obtained for the \caii~H~\&K\ are shown in Fig.~\ref{fig:residual}. Fitting those profiles with a Gaussian gives the residual emission value (the equivalent width of the Gaussian, in \kms).  
\begin{figure*}
\center{\hbox{\includegraphics[scale=0.4]{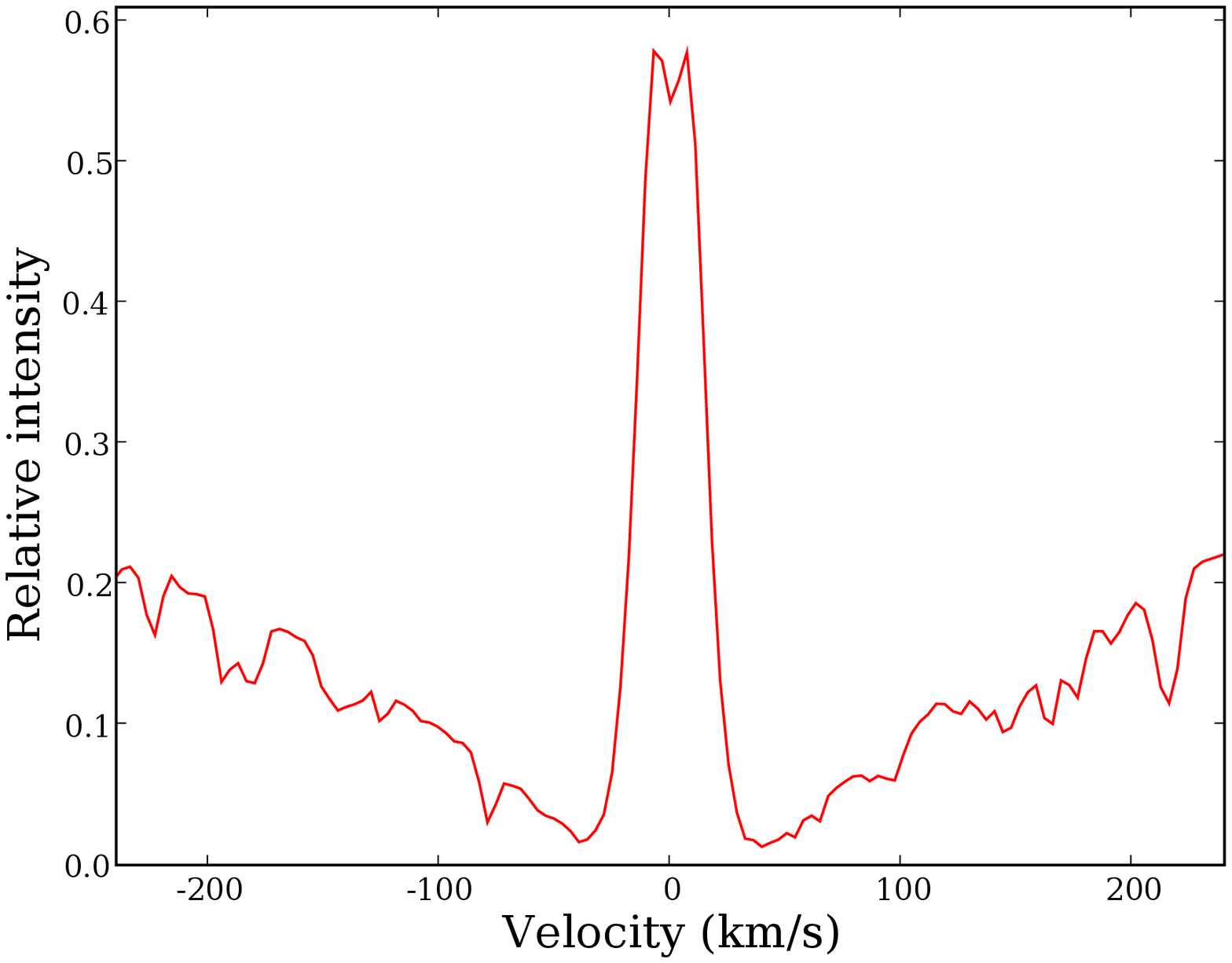}\includegraphics[scale=0.4]{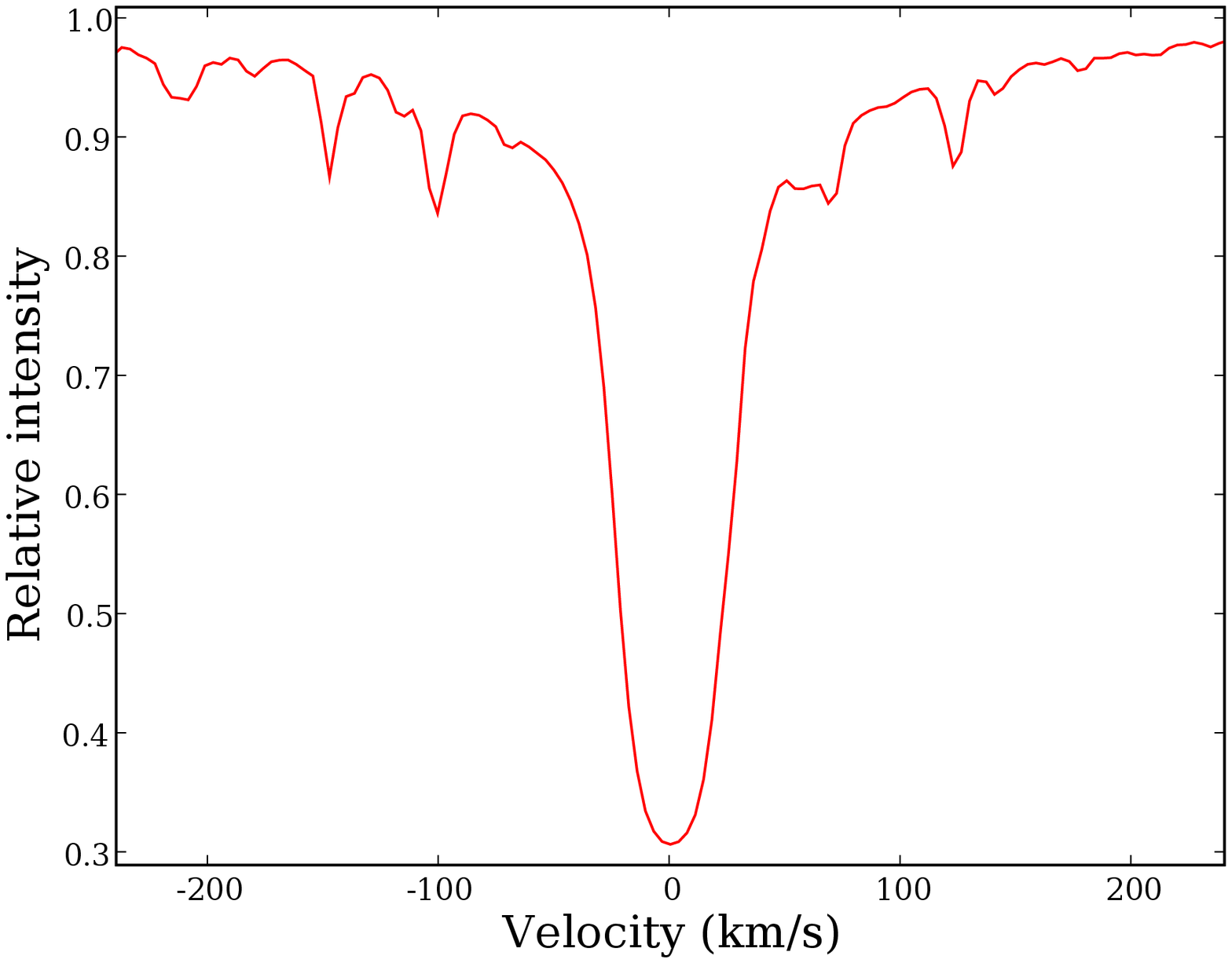}}}
\caption[]{Mean \caii~H~\&~K\ (left panel) and  H$\alpha$ (right panel) profiles for July~2008.}
\label{fig:meanCa}
\end{figure*}

\begin{figure*}
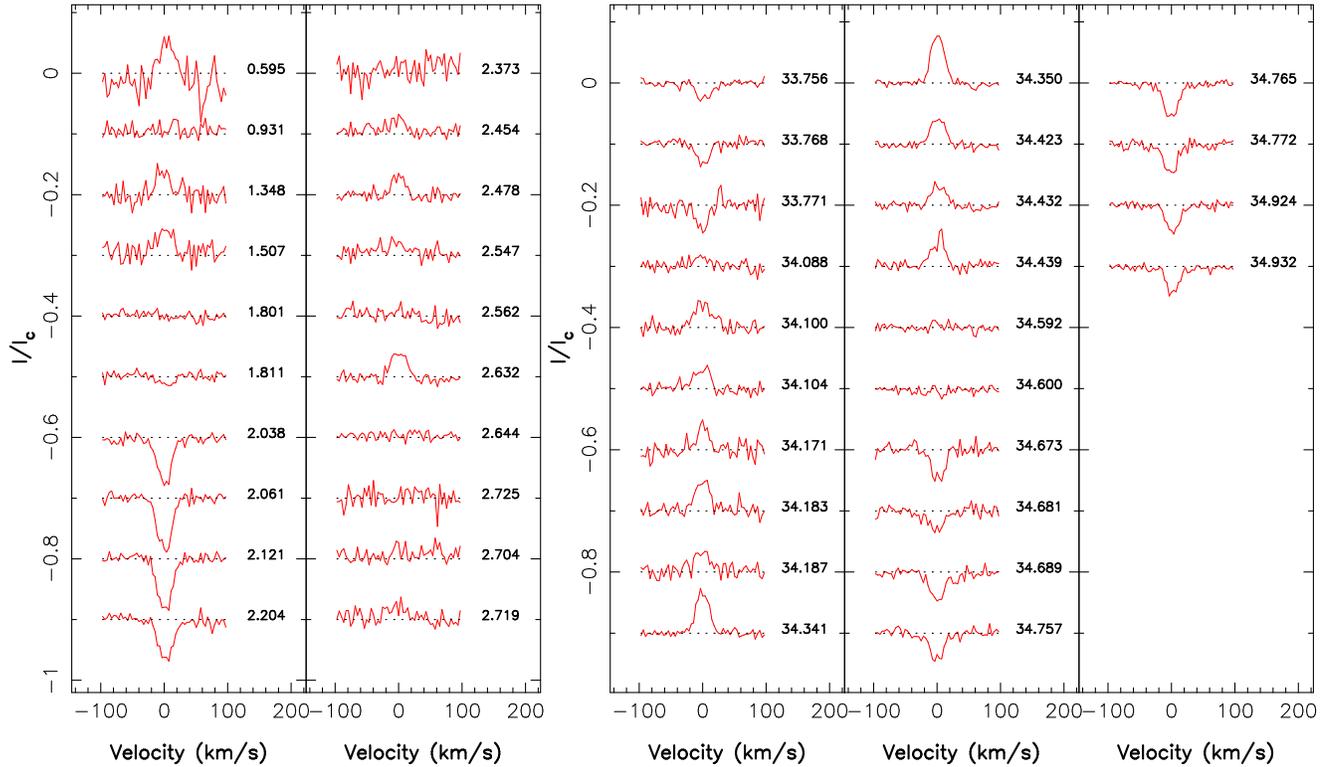

\center{\hbox{\includegraphics[scale=0.55,angle=-90]{Cajun07_article.ps}\includegraphics[scale=0.55,angle=-90]{Cajul08_article.ps}}}
\caption[]{\caii~H~\&~K\ residual profiles for June 2007 (left panel) and July 2008 (right panel). The rotational cycle of the star is mentioned next to each profile. }
\label{fig:residual}
\end{figure*}

\begin{figure*}
\center{\hbox{\includegraphics[scale=0.45]{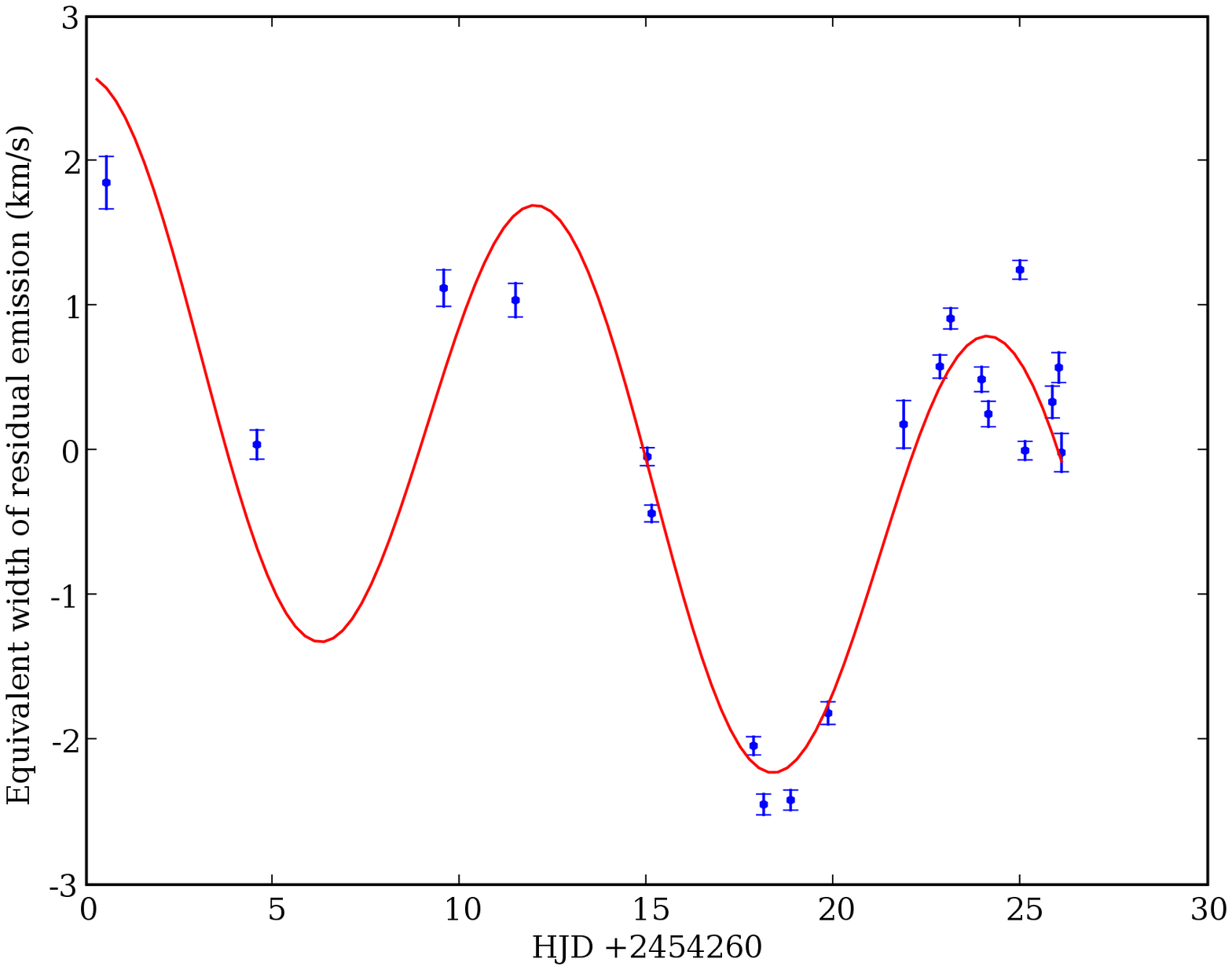}\includegraphics[scale=0.45]{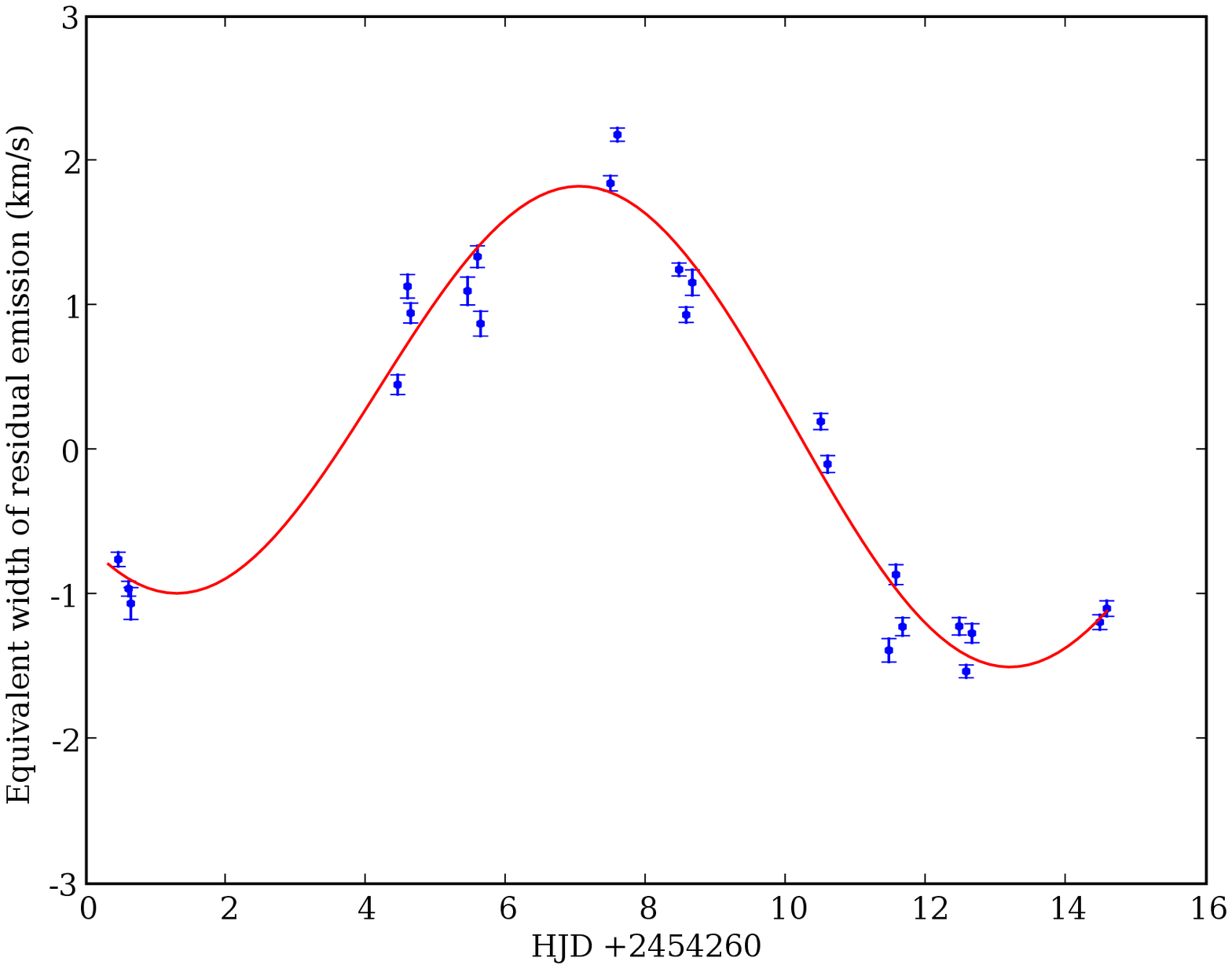}}}
\caption[]{Equivalent width variations in the emission cores of the \caii~H~\&~K\ (after subtracting a mean profile calculated for the run) as a function of the Heliocentric Julian Date for June 2007 (left panel) and July 2008 (right panel). The red curves are the best fit to the data (see text for more details).}
\label{fig:Camodulation}
\end{figure*}
 
We studied the variation of the residual emission for each observing epoch (including 2006). The star exhibits variability on both short (hours) and long (days) time scales. We modelled the longer term variability as the sum of a sine wave (to describe the periodic modulation) and a linear function (to describe the longest term variability).
For a range of period between 2~d and the length of the run, the period for which the \chisqr~of the fit is minimal is the best fit period to the data. 
 The best fit periods are listed in Table \ref{tab:per}. We plotted the \caii~H~\&~K\ residual emission and the best fit solution in Fig. \ref{fig:Camodulation}. For all three epochs, the \caii~H~\&~K\ residual emission is modulated on a time-scale of the equatorial rotation period. For H$\alpha$, the residual emission is modulated on a time-scale of $\sim 13$~d (on average), again close to the rotational period of the equator. H$\alpha$ and \caii~H~\&~K\ residuals are well correlated (see Fig. \ref{fig:correlation}) except at 2 main epochs (2006~August~8-12, 2007~June~9-19) where H$\alpha$ is slightly stronger at a given \caii~H~\&~K\ residual emission. 
\begin{figure*}
\includegraphics[scale=0.45]{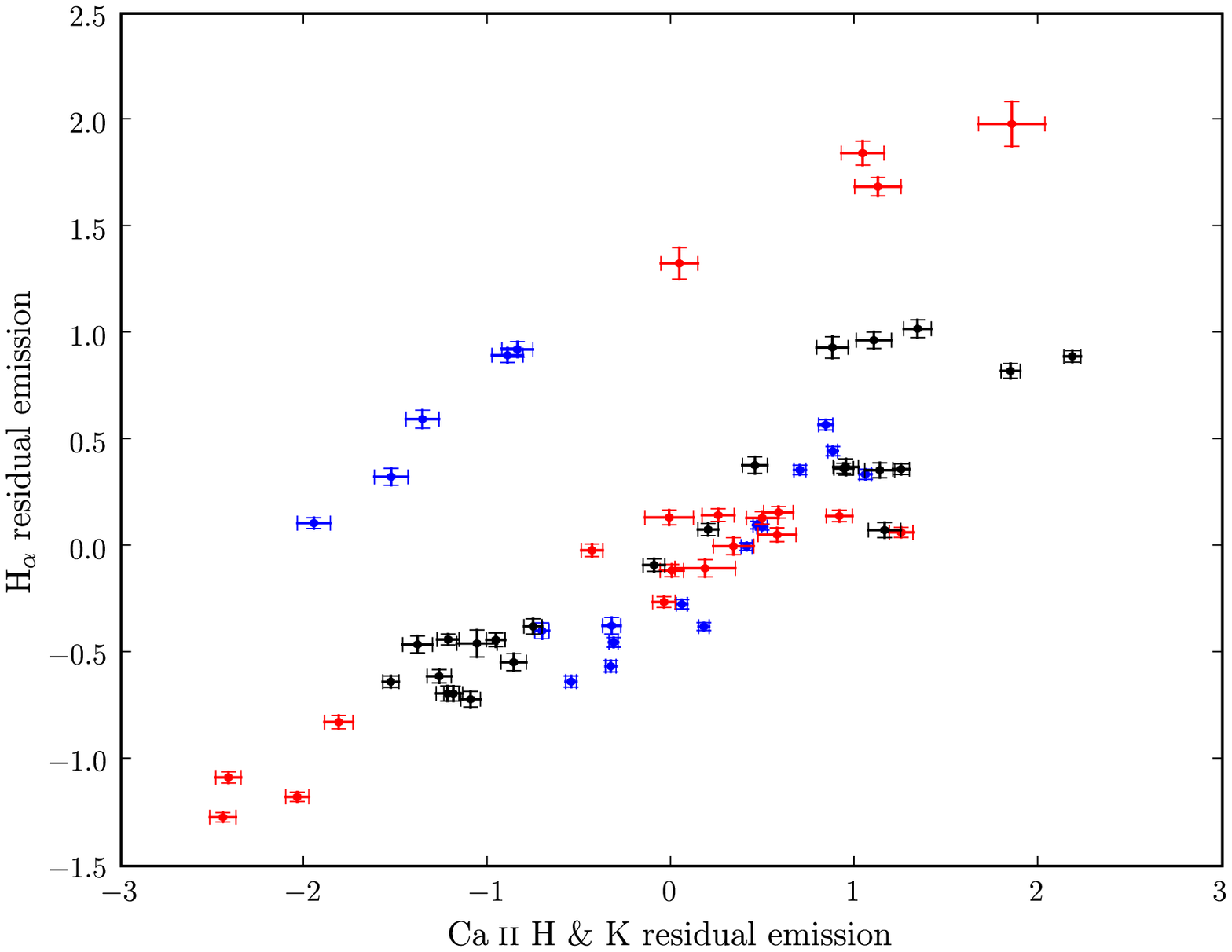}
\caption[]{H$\alpha$ and \caii~H~\&~K\ residual emission for the three observing epochs (blue, red and blacks dots are respectively for 2006, 2007 and 2008 data).}
\label{fig:correlation}
\end{figure*}

\begin{table}
\caption[]{The best fit period for the \caii~H~\&~K\ and the H$\alpha$ residuals in days (using the model described in the text). The error bars correspond to a 3~$\sigma$ error bars on our measurements.}
\begin{tabular}{|c|c|c|}
\hline
Epoch & \caii~Residual  & H$\alpha$~Residual \\
& Period(d)& Period(d)\\
\hline
2006 & $12.15^{+0.3}_{-0.4}$ &$ 11.6^{+0.3}_{-0.2}$ \\
\hline
June 2007 & $12.1^{+2.2}_{-1}$ & $14.4^{+1.5}_{-1.3}$\\
\hline
July 2008 &$ 11.9^{+1.2}_{-0.8}$ & $13.8^{+2.2}_{-2.3}$ \\ 
\hline
\end{tabular}
\label{tab:per}
\end{table}

To look for additional low-amplitude periodic fluctuations, we subtracted the rotational modulation from the data (the best fit solution obtained). In the case of SPI, the emission variability would be modulated by the orbital period or rather by the beat period between the stellar rotation and the planetary orbital period (varying between 2.5 and 2.7~d, depending on the stellar latitude considered). Fig.~\ref{fig:ResRes} shows these residuals as a function of the orbital phase for the \caii~H~\&~K\ . A large dispersion in the residuals can be seen. In particular, we can see that the residual emission is highly dispersed at any given orbital phase range, directly reflecting the short term variability of the activity proxies. We therefore suspect that this is mainly caused by intrinsic variations in the activity of the star rather than to SPI, as we would expect enhanced emission variability to be concentrated at specific phases \citep{shk08}.

We then searched for periodic modulation of these residuals for a range of periods by fitting the residuals with a sine wave. For June~2007, we find a period of $3.65^{+0.25}_{-0.7}$~d in H$\alpha$, significant to $3~\sigma$. We do not find a similar period in the \caii~H~\&~K\ residuals. For July~2008, two periods of $2.45^{+0.15}_{-0.25}$~d and $3.5^{+0.2}_{-0.2}$~d were found in H$\alpha$ and a large peak around 2.9~d in the \caii~H~\&~K\ ($2.9^{+0.5}_{-0.6}$~d). Only in July~2008 for the H$\alpha$ emission, one of the two identified periods is roughly compatible with the beat period.

\begin{figure*}
\center{\hbox{\includegraphics[scale=0.45]{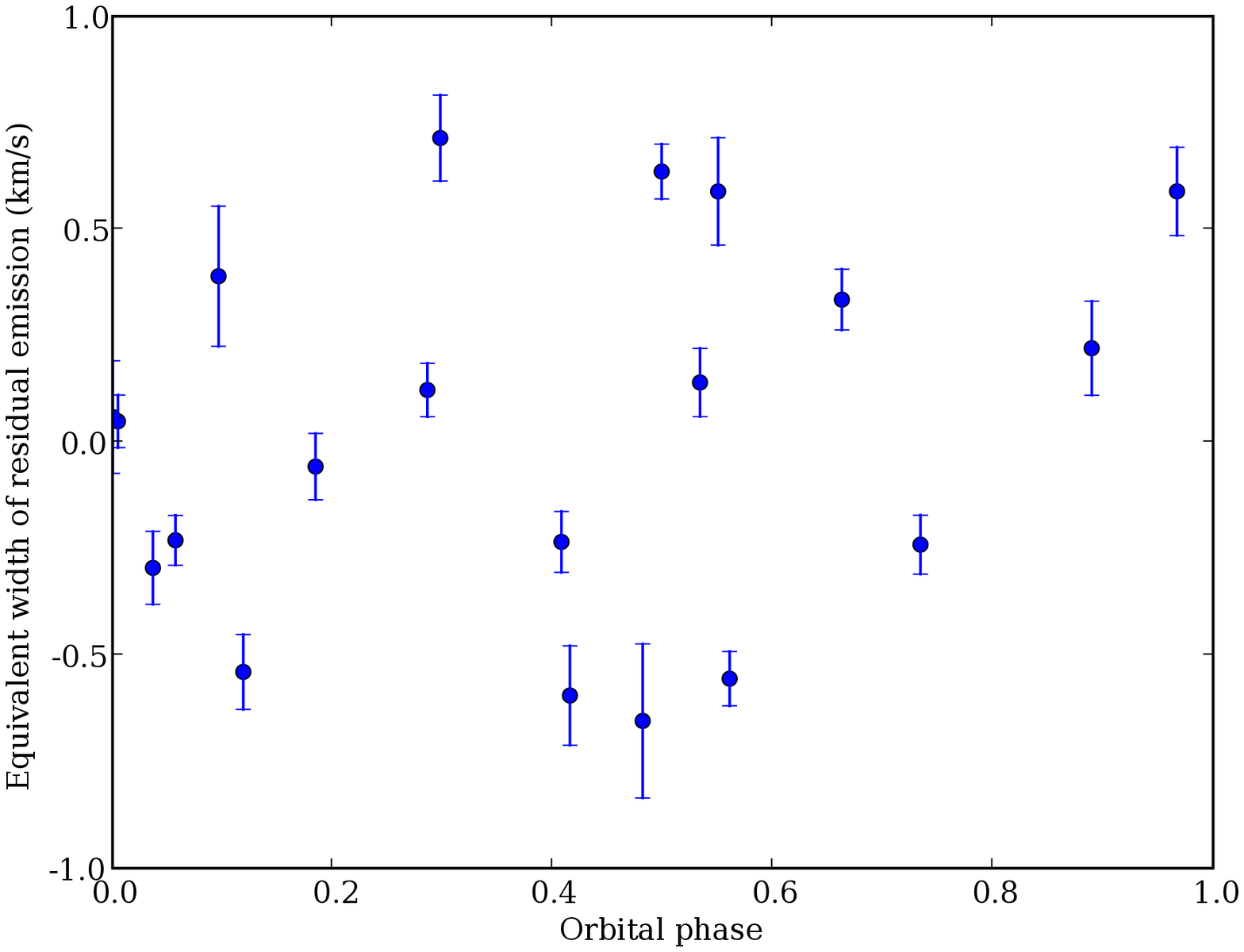}\includegraphics[scale=0.45]{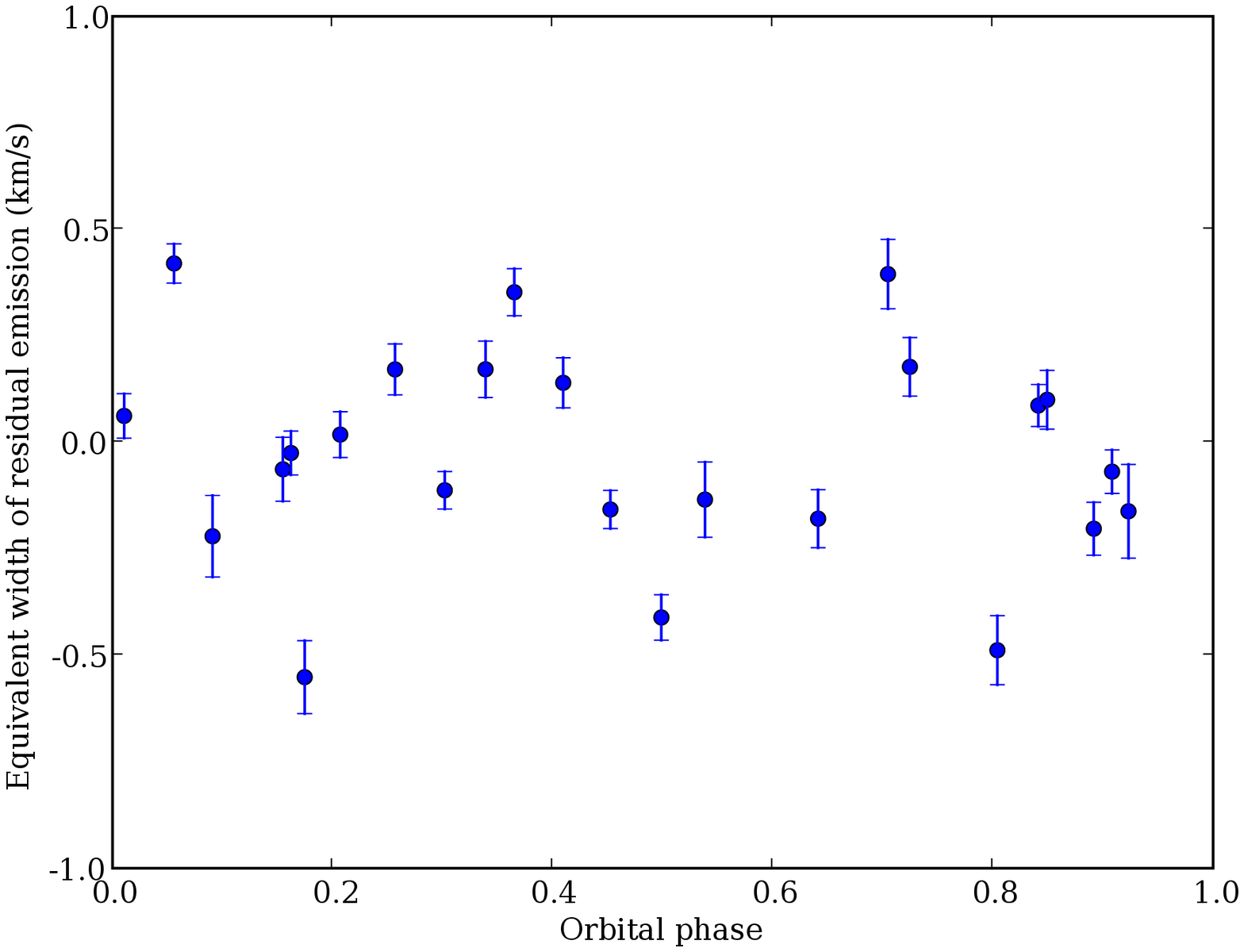}}}
\caption[]{Residuals of the \caii~H~\&~K\ equivalent width after subtracting the rotational modulation, as a function of the planet orbital phase for June 2007 (left panel) and July 2008 (right panel).}
\label{fig:ResRes}
\end{figure*}
 
\section{Extrapolation of the magnetic field}
\label{sec:extrapolation}


From the surface magnetic field, one can then extrapolate the magnetic field in the stellar atmosphere assuming a potential field \citep{jardine02}. 
In the case of SPI, extrapolating the magnetic field lines is a way to understand the environment in which the close-in planet evolves, and to investigate potential  interactions (e.g. reconnection events due to interactions of the stellar and planetary magnetic fields). 

The extrapolation technique was first applied to the solar corona, and then to other stars. As for the Sun, we assume that there is a surface beyond which the field becomes purely radial, named the Source Surface (hereafter SS). All closed field lines, connecting regions on the stellar surface of different magnetic polarity, are inside the SS (the heights of the magnetic loops are smaller than the SS radius). From the extrapolation of the magnetic field, we can calculate the magnetic energy at any given point inside and on the SS. To calculate the magnetic energy outside the SS, we consider that the magnetic flux is constant on radial shells.

In the system HD~189733, the massive planet is at 8.8~$R_{\star}$. The position of the SS is not well known. We first consider that the SS is located at 3.4~$R_{\star}$ (as a lower limit given that the SS radius $R_{\rm SS}\sim3~R_{\odot}$ for the much less active Sun), i.e. inside the planetary orbit. Figure \ref{fig:extrapol} represents an extrapolation of the stellar magnetic field within the source surface for our two epochs of observations. It shows in particular that the magnetic configuration in the stellar corona is complex. We find that the magnetic field at the distance of the planet has a strength that can reach values up to 40~mG (i.e. 4000~nT, see Fig. \ref{fig:energy}). The magnetic energy is null when the planet crosses in front of the intersection of the neutral line with the equator. 

We considered a second case for which the SS is at 5.8~$R_{\star}$, also within the planetary orbit. The mean magnetic field at 5.8~$R_{\star}$~averaged over the longitudes drops by a factor of 2 relative to the previous case. 
\begin{figure*}
\center{\hbox{\hspace{7mm}\includegraphics[scale=0.4]{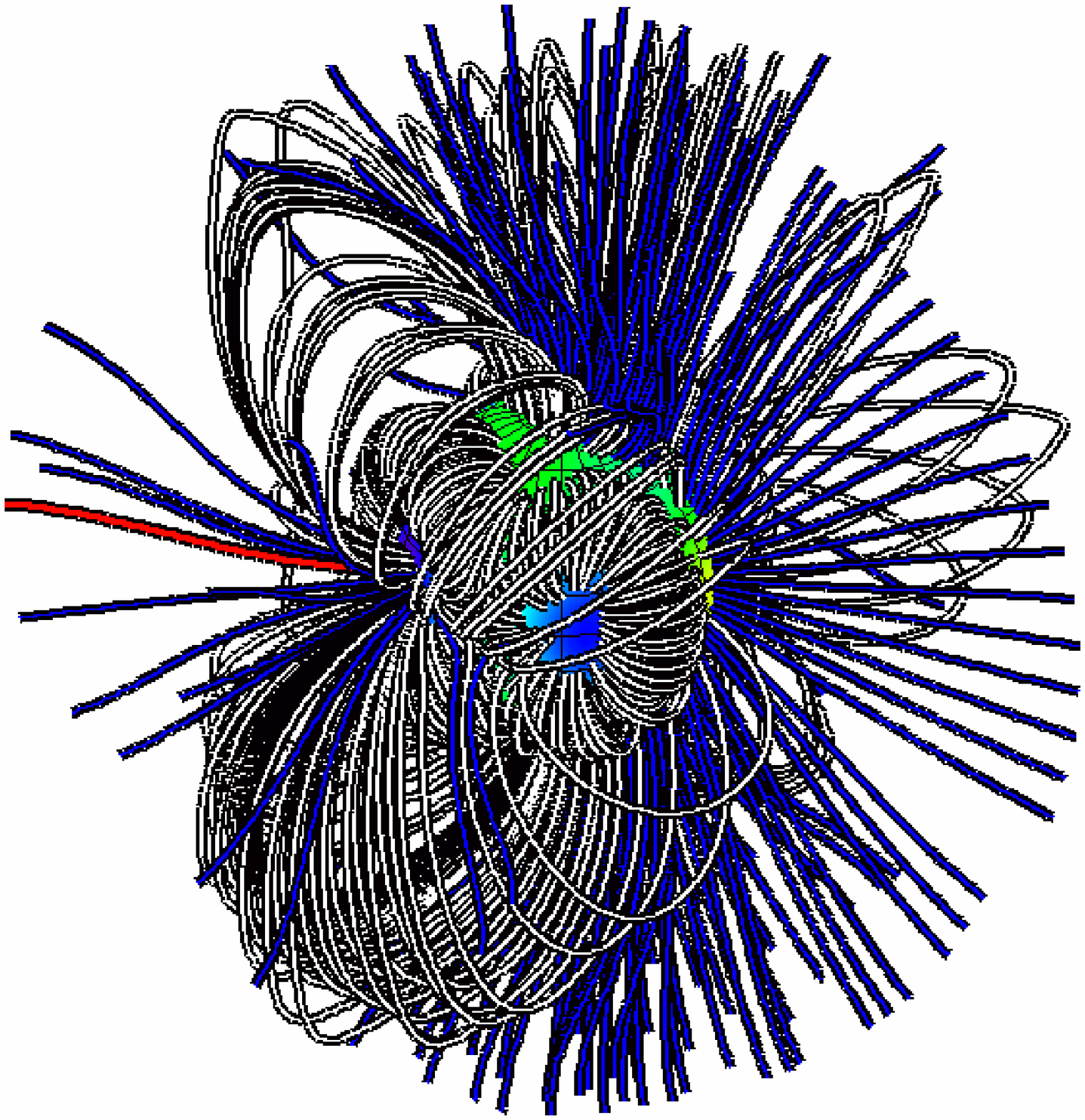}\hspace{7mm}\includegraphics[scale=0.4]{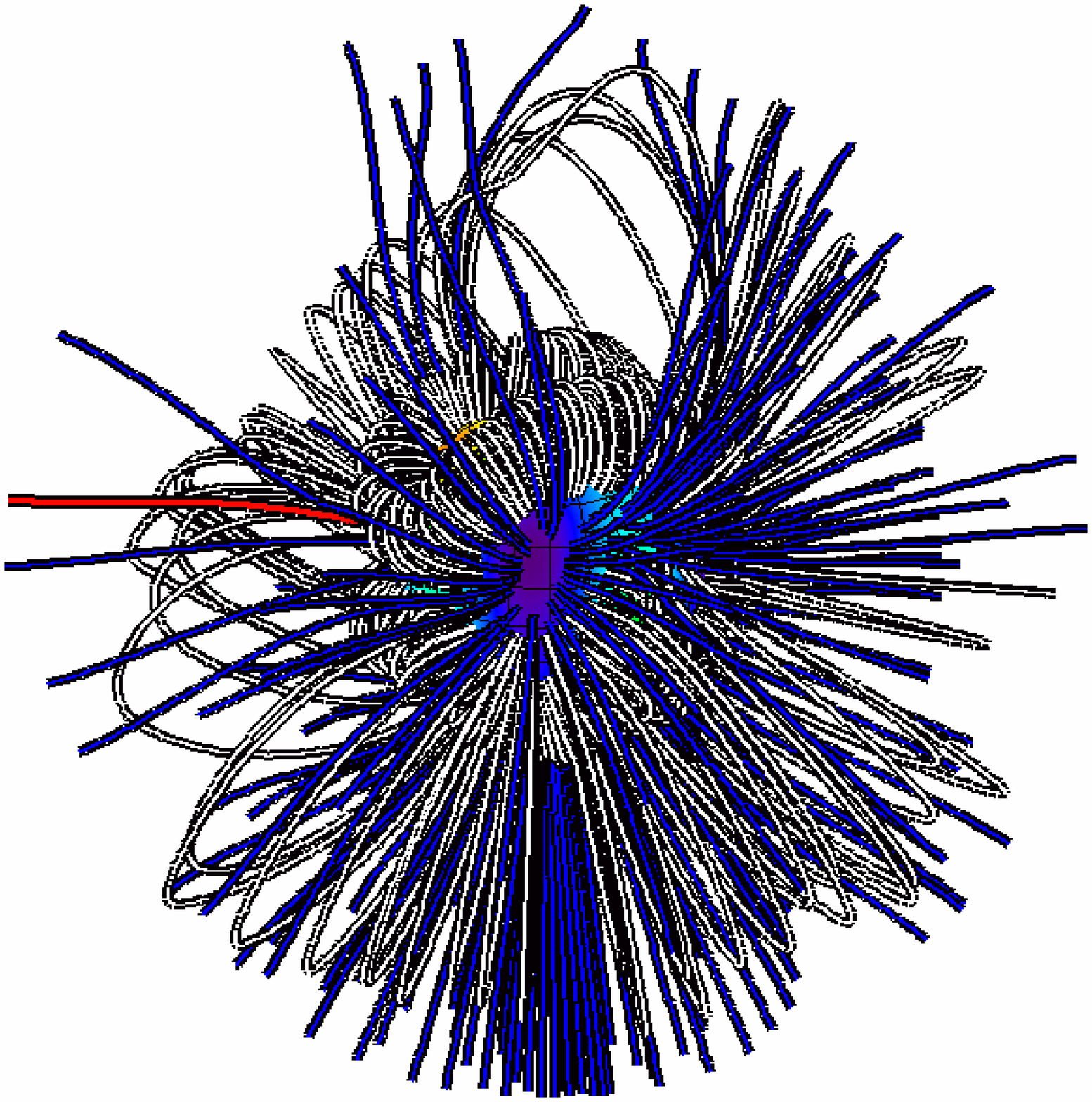}}}

\caption[]{The extrapolated magnetic field of HD~189733 for June 2007 (left) and July 2008 (right). White lines corresponds to the closed magnetic lines, blue ones to the open field lines (reaching the source surface). The red line corresponds to the field line joining the planet to the star at a given (arbitrarily selected) orbital and rotation phase.}
\label{fig:extrapol}

\end{figure*}

\begin{figure*}
\center{\hbox{\hspace{1mm}\includegraphics[scale=0.4]{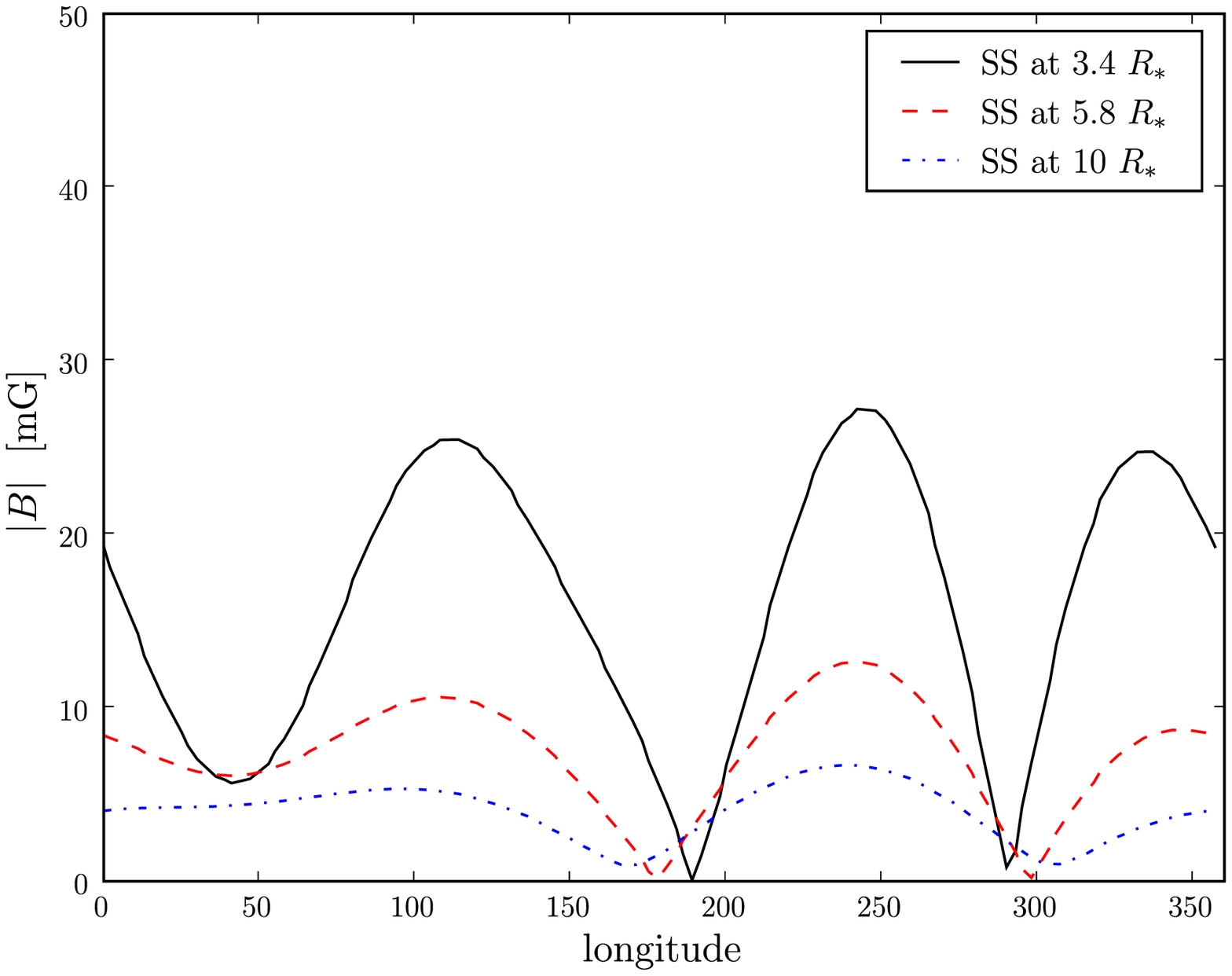}\hspace{1mm}\includegraphics[scale=0.4]{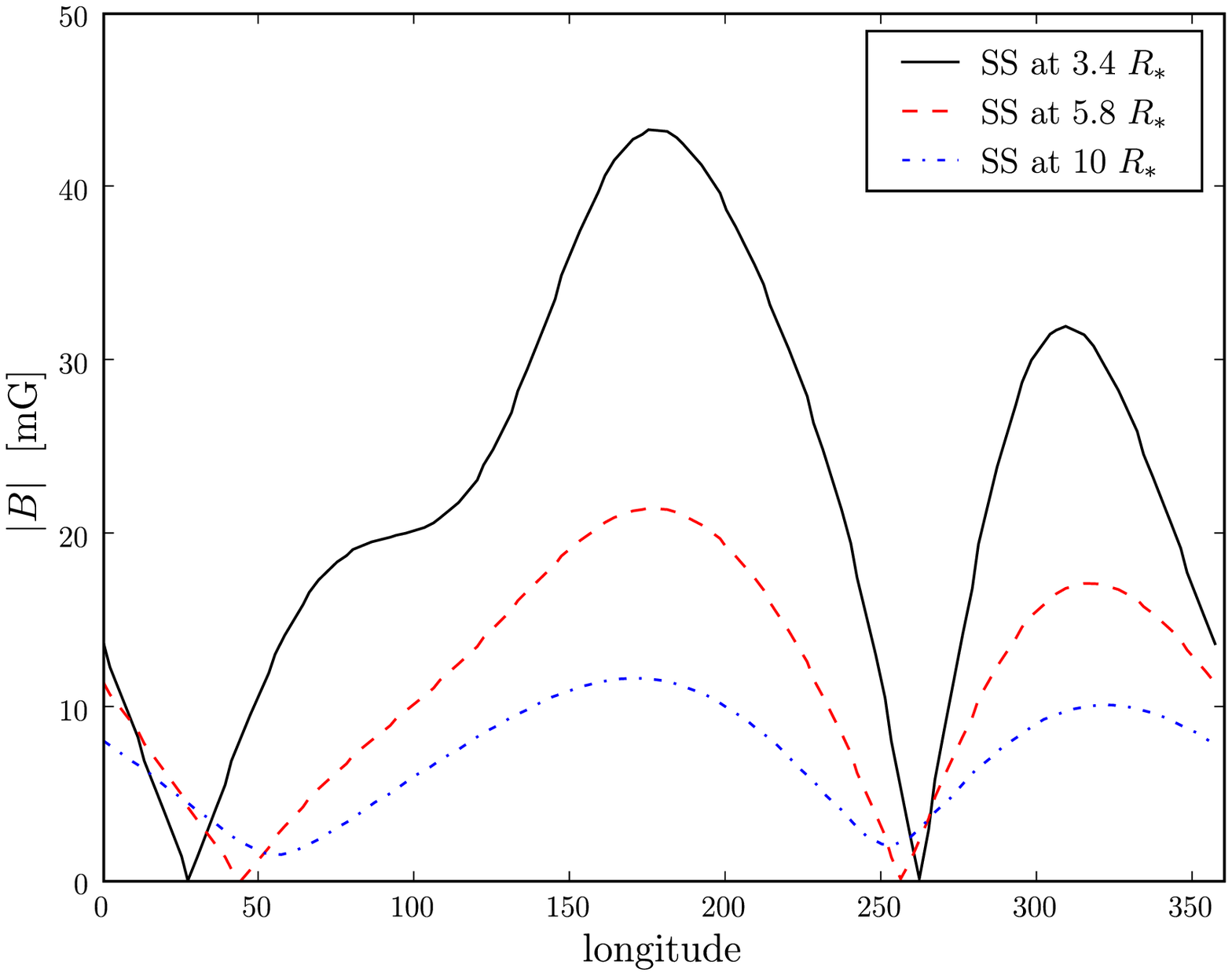}}}

\caption[]{The magnetic field at the distance of the planet as a function of stellar longitude for June 2007 (left) and July 2008 (right) for different positions of the SS (the black, red and blue lines are respectively for a SS placed at 3.4, 5.8 and 10~$R_{\star}$).  }
\label{fig:energy}

\end{figure*}

Finally, we consider cases where the SS is larger than 8.8~$R_{\star}$. When the SS is at 10~$R_{\star}$, we notice that the magnetic field at the planet orbit is smaller than that for a SS at 3.4~$R_{\star}$ by a factor slightly larger than 3 for the mean value (see Fig. \ref{fig:energy}). For a SS beyond 10~$R_{\star}$, the magnetic field at the planet orbit does not change dramatically relative to the previous case. In all these cases, the value of the field is not constant along the planet orbit.

We thus expect average field values of $4-23~\rm mG$ at the distance of the planet.

\section{The expected radio emission}
\label{sec:radio}

Knowing the stellar magnetic field, one can estimate the expected radio emission for HD~189733b. 

In the magnetic energy model \citep{Zarka01}, planetary radio emission is believed to be energized by the Poynting flux transported by the stellar wind. The amount of power emitted by radio waves $P_{\rm {rad}}$ is roughly proportional to the power input $P_{\rm {input}}$ supplied by the stellar wind magnetic energy \citep[see Figure 6][]{Zarka06PSS}. 
Thus, the radio power is given by
\begin{eqnarray}
P_{\rm{rad}} \propto P_{\rm{input}} \propto  v_{\rm{eff}} B_{\perp}^2 R_{\rm{s}}^2
\label{eq:Pin:mag}
\end{eqnarray}
where $v_{\rm{eff}}$ and $B_{\perp}$ are respectively the velocity of the stellar wind and the component of the interplanetary magnetic field (IMF) perpendicular to the stellar wind flow, both in the reference frame of the planet, and $R_{\rm{s}}$ denotes the radius of the planetary magnetosphere (known as magnetospheric standoff distance).

To estimate the radio power, one has to know $B_{\perp}$, $v_{\rm{eff}} $  and $R_{\rm{s}}$. They depend on the stellar magnetic field, rotation and age \citep[see][for more details]{Griessmeier07AA}. When a star is older, its wind strength is weaker. Via the stellar wind strength (velocity and density), the stellar age governs the size of the planetary magnetosphere, i.e.~the cross section on which the planet can intercept the energy flux.

Given the rotation period of 12~d, the stellar age is of $\sim 1.6$~Gyr \citep[based on the formalism presented in][]{Griessmeier07AA}. Using the stellar wind model of \citet{Griessmeier07AA}, we calculate $R_{\rm{s}}$ and $v_{\rm{eff}}$, that are equal respectively to $3.4~ R_{\rm planet}$ and $332~\kms$ (both constant on the planetary orbit).
The magnetic field at the distance of the planet is taken from the results of the extrapolation (see section \ref{sec:extrapolation} and Fig. \ref{fig:energy}). In the particular case where the SS is beyond the planetary orbit, our calculation includes the three components of the field.

In this context, we find average radio fluxes of about 7 to 220~mJy depending on the location of the SS, for a frequency range of 0-6~MHz. Values up to 20~MHz are compatible within the model uncertainties. We also find this radio flux to be time variable (see Fig. \ref{fig:radio}).\footnote{\citet{Griessmeier07AA} have considered different parameters for the system, in particular a stellar age of $\sim 5.2$~Gyr, a rotation period of 28~d, and a radial magnetic field of about $2~\rm{G}$ at the surface of the star yielding a different radio flux.}

\begin{figure}
 \includegraphics[scale=0.6]{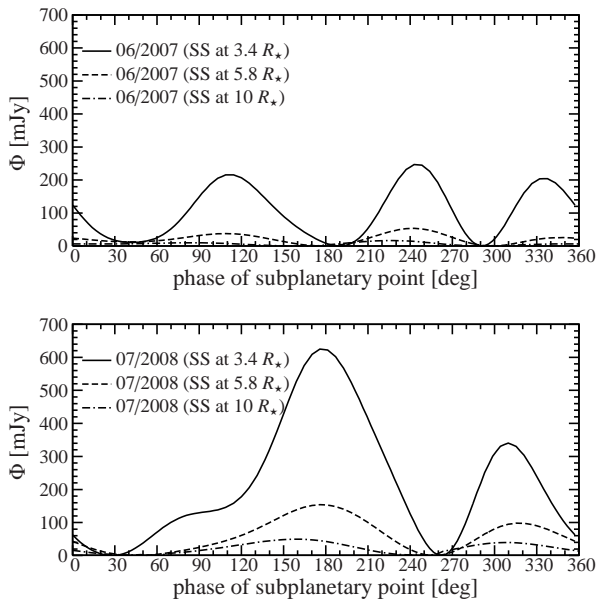}
\caption{
Expected planetary radio flux at Earth as a function of the rotational phase of the 
substellar point, for three different positions of the source surface. 
Top panel: Radio flux for the magnetic configuration of June 2007. 
Bottom panel: Radio flux for the magnetic configuration of July 2008.
\label{fig:radio}}

\end{figure}


\section{Discussion and conclusions}
\label{sec:conclusions}

In this paper, we present a detailed spectropolarimetric study of the star HD~189733, host of a transiting giant planet. The star was observed at two epochs (June~2007 and July~2008). Using Zeeman Doppler Imaging, we reconstructed the magnetic maps of the star. With a strength up to 40~G, the magnetic field is dominated by the toroidal component at both epochs. This component contributes 57\%~and 77\%~to the total energy respectively and is mainly axisymmetric. In contrast, the poloidal component is mainly non-axisymmetric. Its contribution to the total energy drops from 2006 to 2007 and 2008. We will continue monitoring this system to study the magnetic evolution on time-scales longer than 2 years and look for a potential magnetic cycle. 

HD~189733 rotates differentially and has a latitudinal angular rotation shear of $\dom = 0.146 \pm 0.049$~\rpd~; the star has an equatorial period of $11.94\pm0.16$ d and a polar period of $16.53\pm2.43$~d. These values of the equatorial and polar periods bracket all published photometric periods within the error bars. 

The star is an active star, variable on small time-scales. We analyzed the activity residuals in the \caii~H\&K\ and H$\alpha$. These residuals are periodic and modulated with the rotational period of the star. Active regions apparently concentrate around the equator, given the modulation period of 12-13~d. A rotational modulation was also found by \citealt{moutou07}, \citealt{boisse09}, and \citealt{shk08}. We looked for lower amplitude periodic fluctuations, the periods we found are different than the orbital and beat period (2.5-2.7~d) except in one occurrence (H$\alpha$, July 2008) where one of the detected roughly matches the beat period. 

To enlarge our knowledge of this system, we studied the magnetic field in the stellar atmosphere using the extrapolation technique applied to the reconstructed surface magnetic maps. We find that HD~189733 has a complex magnetic topology for both epochs. Depending on where the SS (the surface beyond which the field is purely radial) is, we find that the magnetic field at the distance of the planet ($8.8~R_{\star}$) is variable throughout the orbit, of the order of $4-23~\rm mG$~in average.

We also estimated the radio flux expected from SPI assuming the magnetic scenario model of \cite{Zarka01} and find it to be of the order of $10-220~\rm{mJy}$ on average. We also predict it to be variable with time on a time scale equal to the beat period (contrary to previous published predictions \citealt{Griessmeier07AA}). The radio flux we predict at 0-6~MHz is potentially detectable with LOFAR (see fig. 1-3 in \citealt{Griessmeier07AA}) in the coming years. Published radio observations only report upper limits on the planetary flux at higher frequencies \citep[307-347~MHz,][244~MHz and 614~MHz]{smith09,lecavalier09}~-~providing no constraint on the model discussed in the present paper. The variability of the planetary radio flux with the subplanetary stellar phase will make the distinction between the planetary and stellar radio flux in the observational data more challenging; however, this effect can be used to distinguish between the magnetic energy model and other models of interactions. Our result confirms that a single observation of a star-planet system is not sufficient. Rather, it is important to have multiple observations densely sampling stellar rotation, planetary orbit and beat periods. 

The study of SPI is an ongoing effort. Monitoring stars at different epochs and through multi-wavelength campaigns will help us identify the nature of SPI and the origin of their apparent on-off behavior. Studying stellar magnetic cycles and comparing results for HJ hosting stars with different stellar and planetary parameters will enlarge our understanding of SPI, as well as stellar magnetism and activity in general.   

\section*{Acknowledgments}
This work is based on observations obtained with ESPaDOnS at the Canada-France-Hawaii Telescope (CFHT) and with NARVAL at the T\'elescope Bernard Lyot (TBL). CFHT/ESPaDOnS are operated by the National Research Council of Canada, the Institut National des Sciences de l'Univers of the Centre National de la Recherche Scientifique (INSU/CNRS) of France, and the University of Hawaii, while TBL/NARVAL are operated by INSU/CNRS. We thank the CFHT and TBL staff for their help during the observations. We thank the referee, J.~Landstreet, for his comments on the manuscript. J.-M.~G. was supported by the french national research agency (ANR) within the project with the contract number NT05-1\_42530. 
\bibliography{biblio}
\bibliographystyle{mn2e}

\end{document}